# On the galaxy luminosity function in the central regions of the Coma cluster [*]

A. Biviano[1], F. Durret[1,2], D. Gerbal[1,2], O. Le Fèvre[2], C. Lobo[1,3], A. Mazure[4], and E. Slezak[5]

[1] Institut d'Astrophysique de Paris, CNRS, Université Pierre et Marie Curie, 98bis Bd Arago, F-75014 Paris, France
[2] DAEC, Observatoire de Paris, Université Paris VII, CNRS (UA 173), F-92195 Meudon Cedex, France
[3] Centro de Astrofísica da Universidade do Porto, Rua do Campo Alegre 823, 4100 Porto, Portugal
[4] LAS, Traverse du Siphon, Les Trois Lucs, B.P. 8, F-13376 Marseille Cedex, France
[5] Observatoire de la Côte d'Azur, B.P. 229, F-06304 Nice Cedex 4, France

**Abstract.** We have obtained new redshifts for 265 objects in the central $48 \times 25$ arcmin$^2$ region of the Coma cluster. When supplemented with literature data, our redshift sample is 95 % complete up to a magnitude $b_{26.5}=18.0$ (the magnitudes are taken from the photometric sample of Godwin et al. 1983). Using redshift-confirmed membership for 205 galaxies, and the location in the colour-magnitude diagram for another 91 galaxies, we have built a sample of cluster members which is complete up to $b_{26.5}=20.0$.

We show that the Coma cluster luminosity function cannot be adequately fitted by a single Schechter (1976) function, because of a dip in the magnitude distribution at $b_{26.5} \sim 17$. The superposition of an Erlang (or a Gauss) and a Schechter function provides a significantly better fit.

We compare the luminosity function of Coma to those of other clusters, and of the field. Luminosity functions for rich clusters look similar, with a maximum at $M_b \simeq -19.5 + 5 \times \log h_{50}$, while the Virgo and the field luminosity functions show a nearly monotonic behaviour. These differences may be produced by physical processes related to the environment which affect the luminosities of a certain class of cluster galaxies.

## 1. Introduction

The study of luminosity functions (hereafter LFs) is a powerful means to constrain theories of galaxy formation and evolution, and allows to estimate the baryon density in the Universe as provided by galaxies. After pioneering work due to Hubble & Humason (1931) and Zwicky (1942), Schechter (1976) has proposed an analytical form now widely used, although recent investigations by Katgert (private communication) and Metcalfe et al. (1994) tend to favour the analytical form suggested by Abell (1962). The universality of the galaxy LF has been a controversial issue for many years (see, e.g., Oegerle & Hoessel 1989; Lugger 1989; and references therein), and many investigations have been devoted to galaxy LFs in cluster environments, since a high galaxy density may favour particular conditions of evolution, which could be reflected in modifications of the galaxy LF. With the determination of the LFs of Virgo cluster galaxies, Sandage et al. (1985) have made a distinction between galaxies of different morphological types and surface brightenesses. In particular, they have shown that the bright galaxy LFs are bounded both at the bright and faint limits, while dwarf galaxies follow an ever-increasing distribution at the faint end.

The LF of the Coma cluster has been investigated by many authors, in particular by Rood (1969), Godwin & Peach (1977) and Godwin et al. (1983, hereafter GMP). GMP have provided a photometric sample of galaxies complete to $b_{26.5} = 20.0$ in a 2.63 degree$^2$ field centered on the Coma cluster. Rood (1969) noted a peak in the Coma LF around $V_{26} \simeq 14.7$, later confirmed by Godwin & Peach (1977). Thompson & Gregory (1993, hereafter TG) found a much less pronounced maximum at $b_{26.5} \simeq 16.5$, and fitted the bright galaxy LF by a Gauss distribution, as suggested by Sandage et al. (1985) for the Virgo cluster. However, all these investigations lacked the spectroscopic information needed to assign cluster membership reliably; even TG have complete redshift information only for galaxies to a limiting photographic magnitude of 15.7, i.e. one magnitude off the apparent peak in the LF. Our new data-set allows us to constrain the bright part of the Coma only. A first presentation of our results can be found in Biviano et al. (1994).

## 2. Observations

Using the MOS-SIS spectrograph (Le Fèvre et al. 1994) at the Canada-France-Hawaii Telescope during a four night observing run on May 23–27, 1993, we observed a total field of about $48 \times 25$ arcmin$^2$ centered on the two brightest central galaxies of Coma (NGC 4874 and NGC 4889) – at the distance of the Coma cluster, 10 arcmin correspond to 0.4 $h_{50}^{-1}$ Mpc. The O300 grism was used to provide a spectral resolution of 17 Å and a spectral coverage of at least 4000 to 7500 Å. Spectra have been reduced using the MULTIRED software developed by one of us (O. Le Fèvre) in the IRAF environment. The total number of reliable redshifts obtained is 265. The final accuracy in the velocity measurements was estimated from repeated measurements on the same object in different masks, and comparison with existing measurements in the literature : we found for 86 objects that the rms velocity error is 96 km s$^{-1}$. Our data and a more complete description of the quality of our redshifts together with the observing log, will be submitted to A&AS in the next future.

by $-1200"\leq x \leq 1700"$, $-750"\leq y \leq 750"$, with respect to the center defined in GMP ($\alpha_{1950}=12^h57^m.3$; $\delta_{1950}=28°14'.4$), where our observations have been taken. After inclusion of the velocities from the literature (Mazure et al. 1988; Caldwell et al. 1993) we have a sample of 277 objects with available velocities in the above-defined central region, up to the above-defined magnitude limit. From the histogram displayed in Fig. 1, it can be seen that this redshift sample is complete up to a magnitude $b = 17.0$, and almost complete (95 %) up to $b = 18.0$.

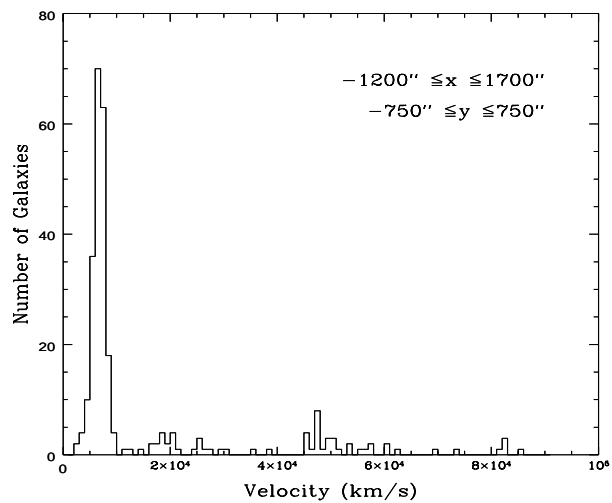

**Fig. 2.** Histogram of velocities for galaxies in the Coma cluster central region.

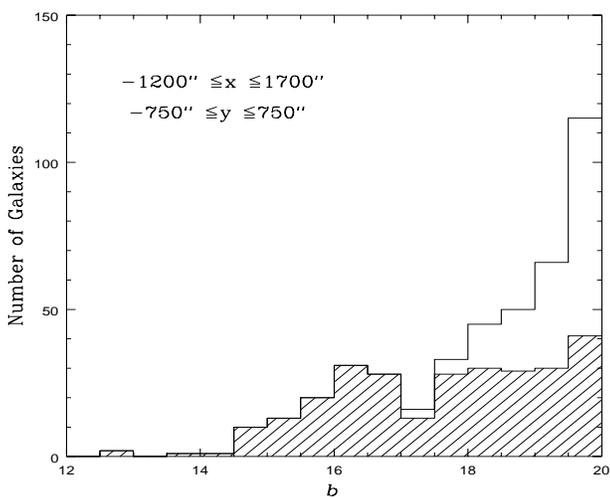

**Fig. 1.** Magnitude histogram for galaxies in the GMP catalogue, limited to the central region of $48 \times 25$ arcmin$^2$. The hatched histogram superimposed corresponds to galaxies for which we have redshifts.

## 3. Definition of the cluster sample

In order to derive the LF of the Coma cluster, we use the $b_{26.5}$ magnitudes of the GMP catalogue (denoted simply as "b" hereafter), up to its completeness limit,

We have defined the sample of cluster members as the sum of two samples :
- Sample 1 : galaxies with membership based on their redshift. In Fig. 2 we show the histogram of velocities for galaxies located in the above defined central region, from 1000 to 100000 km s$^{-1}$. There is a main peak which appears, yet a more refined analysis is needed to define the cluster. To this purpose, we have examined the plot of velocities vs. clustercentric distances (see Fig. 3), and have selected as cluster members the galaxies with velocities in the range 3000–10000 km s$^{-1}$. Similar ranges are obtained by the application of usual methods for the definition of cluster membership in the velocity space (see, e.g., Girardi et al. 1993 and references therein). This sample contains 205 galaxies.
- Sample 2 : galaxies without available redshift, with membership established using their location in the colour–magnitude (CM) band defined by Mazure et al. (1988). The $(b - r)$ colours are taken from GMP. This sample contains 91 galaxies.

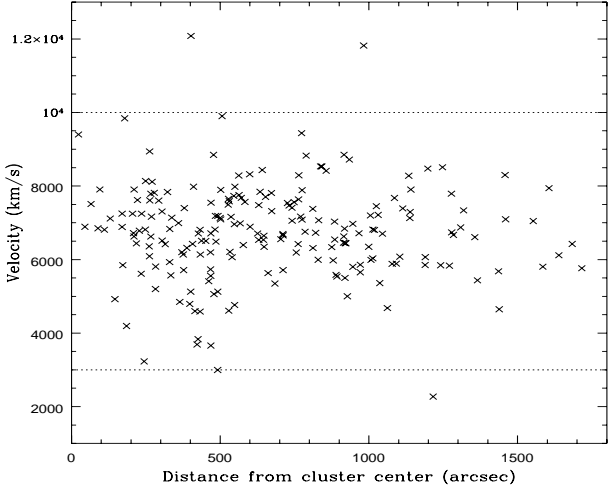

**Fig. 3.** Galaxy velocities vs. distances to the Coma cluster center; the two dotted lines indicate the velocity range within which we consider a galaxy to be a member of the cluster.

are therefore confident that our criteria for cluster membership are acceptable even in the magnitude range where we do not have completeness in redshift.

**Table 1.** Colour–Magnitude Relation

| mag range | $\frac{(\text{\# gal.} \in \text{Sample 1})}{(\text{\# gal. with vel.})}$ | $\frac{(\text{\# gal.} \in \text{CM band})}{(\text{\# gal. with vel.})}$ |
|---|---|---|
| $b \leq 16$ | 1.00 | 0.98 |
| $16 < b \leq 17$ | 0.98 | 0.93 |
| $17 < b \leq 18$ | 0.92 | 0.90 |
| $18 < b \leq 19$ | 0.72 | 0.83 |
| $19 < b \leq 20$ | 0.25 | 0.41 |
| whole | 0.75 | 0.79 |

The galaxies with redshift can be used to test the efficiency of the CM band criterion in identifying cluster members. In Table 1, we list in col.(2) the ratio of the number of galaxies with velocities in the range 3000–10000 km s$^{-1}$ to the total number of galaxies with measured velocities, and in col.(3) the ratio of the number of galaxies with measured velocities which fall inside the CM band to the total number of galaxies with measured velocities, in several magnitude intervals. By comparing the values in col.(2) and col.(3) it can be seen that the CM band selects approximately the correct fraction of galaxy members in every magnitude bin, at least up to a magnitude limit $b \sim 19$ (see also Fig. 4). Although we cannot exclude the presence of a population of blue dwarf galaxies belonging to the cluster, the fraction of these galaxies should be small (TG).

Our final sample of Coma cluster members therefore contains a total of 296 galaxies. In the same region, and up to the same magnitude limit, there are 431 galaxies in the GMP catalogue; 135 of these are therefore likely to be field galaxies. This estimate compares fairly well with the number of field galaxies predicted by the relation of Colless (1989) : 99±50.

We have obtained yet another statistical estimate of the number of interlopers in this field, using the following procedure. First, we have computed the ratios of the number of galaxies with redshifts outside Coma to the total number of galaxies with available redshifts, in bins of half a magnitude. We have multiplied these ratios by the number of galaxies without measured redshifts and found a total number of 89 objects. Second, we have added to this estimate the number of galaxies with redshifts outside Coma (77). The resulting number of interlopers is 166,

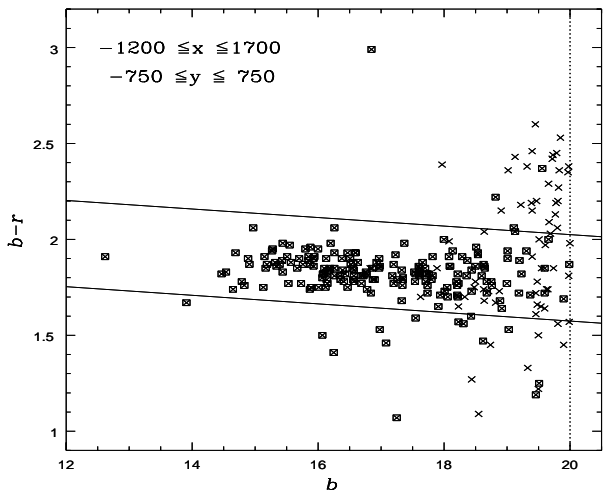

**Fig. 4.** GMP $(b - r)$ colours vs. $b$ magnitudes for galaxies in the Coma cluster region; galaxies with available redshift are denoted by x's, and those belonging to the 3000–10000 km s$^{-1}$ interval are denoted by open squares. The two continuous lines denote the colour–magnitude band defined in Mazure et al. (1988). The dotted line shows the completeness magnitude limit.

## 4. Luminosity functions : fitting results

We used a Maximum-Likelihood (M-L) method (e.g. Malumuth & Kriss 1986) to fit the differential LF of cluster galaxies with the following functions and/or their com-

and 3) a Gamma distribution (also called an Erlang function) :

$$S(b) = K_S \ 10^{0.4(\alpha+1)(b^*-b)} \exp[-10^{0.4(b^*-b)}] \quad (1)$$

$$G(b) = K_G \ \exp[-(b-\mu_b)^2/(2\sigma_b^2)] \quad (2)$$

$$E(b) = K_E(b_E - b)^{a-1} \exp[\lambda(b-b_E)] \quad (3)$$

Two of the most important features of the (M-L) method are that :

- it is well adapted to small samples
- it does not require a binning of the data, at least with the technique proposed by Malumuth & Kriss (1986)

We use various methods to test the quality of our fits :

- the M-L-Ratio test (see e.g. Meyer 1975);
- the Kolmogorov-Smirnov (KS) test;
- the $\chi^2$ test.

In order to perform the last test, we have binned our data. Notice that for some bins the number of galaxies in each bin is very small, so that the statistical significance is dubious (therefore justifying the use of a M-L method in the fitting procedure). Our $\chi^2$ must therefore be considered more a "weighted distance" than a real $\chi^2$. The KS test is a non-parametric test and as such it does not require binning of the data. While the KS and the $\chi^2$ tests give the probability of rejection of the fit, the M-L-Ratio test can only be used to compare the relative qualities of two fits. The simultaneous use of these three tests strengthtens our confidence on the quality of our fits. Results are given in Table 2, where we list the functions used in the fits, the magnitude ranges over which we fit the data, the values of the function parameters, and the probabilities that the fits are rejected according to the KS and $\chi^2$ tests.

In Fig. 5 we show the magnitude histogram for Coma cluster members, and the best fit Schechter function (we stress that although we choose to bin the data for graphical purposes, our fitting procedure requires no binning; similarly, the Poissonian error bars displayed in the figures have not been used in the fitting). As can be seen from the figure, a Schechter function alone does not fit well our data (see also Table 2). This is not surprising given the significance of the dip at $b \sim 17$. A Gauss function is a good fit to the brightest part of the magnitude distribution; a combination of a Gauss and a Schechter functions gives a good fit to the whole magnitude distribution (see Fig. 6 and Table 2). The M-L-Ratio test shows that a Gauss plus a Schechter function give a significant better fit to the data than a single Schechter function, at a 99.55% significance level. Therefore, a Gauss function seems to represent well the magnitude distribution of the bright galaxies in the Coma cluster, while a Schechter function may still represent the faint part of the Coma LF.

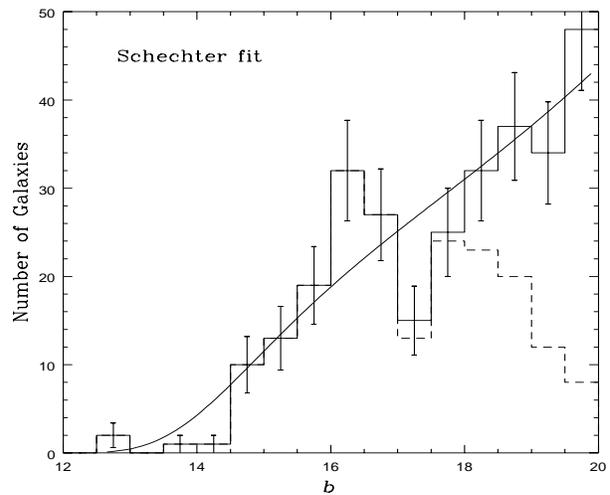

**Fig. 5.** Differential LF for galaxies belonging to Coma; the dashed line histogram is for galaxies with membership established on the basis of their redshifts (Sample 1). Poissonian error bars are displayed. A fit with a Schechter function is shown. The parameters of the function have not been evaluated to fit this particular histogram but are determined by the M-L method, using all 296 data-points.

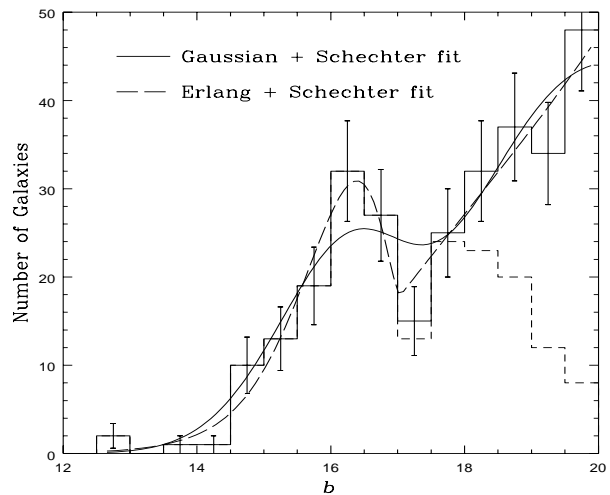

**Fig. 6.** Differential LF for galaxies belonging to Coma; the same histograms as in Fig. 5 are shown. Two fits, one with a Schechter and a Gauss function (continuous line), the other with a Schechter and an Erlang function (long-dashed line), are shown. The parameters of the functions have not been evaluated to fit this particular histogram but are determined by the M-L method, using all 296 data-points.

**Table 2.** Different fits to the luminosity function

| Function | mag range of the fit | $\mu_b$ | $\sigma_b$ | $b^*$ | $\alpha$ | $b_E$ | a | $\lambda$ | Rejection probability KS / $\chi^2$ |
|---|---|---|---|---|---|---|---|---|---|
| S     | $b \leq 20.0$            |            |           | 14.4±0.3 | -1.17±0.06 |          |         |         | 0.525 / 0.997 |
| G     | $b \leq 17.25$           | 16.4±0.2   | 1.1±0.2   |          |            |          |         |         | 0.253 / 0.506 |
| G + S | $b \leq 20.0$            | 16.5±0.3   | 1.2±0.2   | 18.6±0.6 | -0.8 ±0.4  |          |         |         | 0.024 / 0.384 |
| E + S | $b \leq 20.0$            |            |           | 15.9±1.3 | -1.2±0.2   | 17.0±0.1 | 2.6±0.9 | 2.0±0.6 | 0.001 / 0.262 |
| S     | $13.0 \leq b \leq 20.0$  |            |           | 15.1±0.2 | -1.07±0.07 |          |         |         | 0.735 / 0.873 |
| G     | $13.0 \leq b \leq 17.25$ | 16.3±0.1   | 0.9±0.1   |          |            |          |         |         | 0.004 / 0.759 |
| G + S | $13.0 \leq b \leq 20.0$  | 16.0±0.4   | 0.8±0.1   | 17.2±2.5 | -1.1 ±0.5  |          |         |         | 0.004 / 0.589 |
| E + S | $13.0 \leq b \leq 20.0$  |            |           | 15.1±0.6 | -1.3±0.1   | 17.1±0.4 | 4.2±2.4 | 3.4±1.2 | 0.005 / 0.558 |

**Notes to Table 1.** G=Gauss, S=Schechter, E=Erlang

An even better fit to the bright part of the LF can be obtained using an Erlang instead of a Gauss function (see Fig. 6 and Table 2). This happens because of the asymmetry in the bright part of the LF, which cannot be fitted by a Gauss function. However, the M-L-Ratio test shows that the significance level of the difference between the two fits is only 88.38%.

The exclusion of the two brightest cluster members (BCMs) improves most fits, with the notable exception of the Erlang fit. Excluding the two BCMs has the effect of reducing the value of the Gaussian parameter $\sigma_b$ from 1.2±0.2 to 0.8±0.1. The main merit of the Erlang function is its ability to fit the very bright part of the LF. This is due to the mathematical properties of the Erlang function, which is an asymmetric function defined on $]-\infty, \text{limit}]$, and decreasing slowly towards zero at $-\infty$.

We point out that the values of $\mu_b$ and $b_E$, which characterize respectively the maximum and the dip in the bright part of the LF, are quite well constrained; this is not the case for the values of the Schechter function parameters (compare the results for the different fits in Table 2).

## 5. Luminosity functions in different environments

### 5.1. Coma and other rich clusters

We have confirmed that the bright part of the Coma LF has a maximum and therefore cannot be fitted by a monotonically rising function. We have found that a good analytical representation of the Coma LF is given by a combination of an Erlang (or Gauss) function plus a Schechter function. The LF peaks at $b \simeq 16.3$ and has a dip at $b \simeq 17$, corresponding respectively to absolute magnitudes of $M_b \simeq -19.5 + 5 \times \log h_{50}$ and $M_b \simeq -18.8 + 5 \times \log h_{50}$ (we use a mean cluster velocity of 6800 km s$^{-1}$ and a cosmological deceleration parameter q$_0$=1/2, yielding a distance-modulus for the Coma cluster of $35.68 - 5 \times \log h_{50}$ mag, a K-correction of 0.12 mag, and a null galactic absorption).

The particular shape of the Coma LF has been noted by many authors (see §1), but these previous investigations lacked extensive redshift information, unlike our catalogue which is 95 % redshift complete up to $b = 18$, i.e. almost two magnitudes fainter than the maximum in the bright part of the LF. TG have recently performed a detailed investigation of the Coma LF, using a Gauss and a Schechter function to parametrize the LF. Unfortunately, it is difficult to ascertain the consistency of our results with theirs, since the values they find for the function parameters vary from fit to fit (possibly because the fitting technique they use, a standard nonlinear least-squares, is not well adapted to poor statistics).

A similar shape of the LF has been reported in Shapley 8 (see Fig. 7 in Metcalfe et al. 1994), and in Abell 963 (Driver et al. 1994). We have fitted their data with a combination of a Gauss and a Schechter function, and found that the parameters of the Gaussians are very similar to those of Coma; in Shapley 8 we find $\mu_{M_b} \simeq -19.6 + 5 \times \log h_{50}$, and $\sigma_{M_b} \simeq 1.0$; in Abell 963 we find $\mu_{M_b} \simeq -19.3 + 5 \times \log h_{50}$ (using an average colour $(b-r) = 1.8$ for the galaxies in this cluster), and $\sigma_{M_b} \simeq 0.9$.

### 5.2. Virgo and the field

In order to compare our results to those found for the Virgo cluster, we allow for a difference in distance-moduli between Coma and Virgo of 3.68 mag, a K-correction difference of 0.10 mag, and an additional 0.3 mag shift for converting isophotal GMP magnitudes into total magnitudes (TG).

The LF of the Virgo cluster galaxies has been studied in detail by Sandage et al. (1985). As shown by these represented by a function bounded both at the bright and faint ends, and looks similar to the bright galaxy LF in Coma (but see Capaccioli et al. 1992). However, the total LF of Virgo bright and dwarf galaxies looks smooth and monotonic : there is no dip in the total LF, neither for galaxies of all morphological types, nor for early type galaxies only. This is due to the fact that in Virgo, there is quite a large magnitude superposition between the LFs of bright and dwarf galaxies.

Let us have a closer look at the difference between the Coma and Virgo LFs. A cluster LF arises from the superposition of the LFs of Es, S0s, and spirals, and of the dwarf LF. The Gaussian fitted by Sandage et al. (1985) to the Virgo Es and S0s, has a significantly fainter mean and larger dispersion than the Gauss LF in Coma (the mean is $\approx 0.7$ mag fainter, and the dispersion is 1.7 mag, instead of 1.1±0.2 in Coma). The Virgo spiral LF is fitted by a Gaussian with a yet fainter mean and comparable dispersion (1.5 mag); in our sample of bright galaxies ($b \leq 17$) the fraction of spirals is less than 10 %, so the Gaussian part of the Coma LF is populated almost exclusively by Es and S0s. Finally, the LFs of the dwarf galaxies in Virgo and Coma start at roughly the same absolute magnitude, which correponds to the dip in the Coma total LF. Being broad and peaked at a (relatively) faint magnitude, the bright galaxy LF in Virgo largely overlap the bright end of the dwarf galaxy LF, so that no dip is produced in the total LF.

The field LFs for different morphological types have been recently obtained by Marzke et al. (1994), who fitted them with single Schechter functions over the whole magnitude range. Whatever the galaxy morphological type, there is no dip in the bright part of their LFs. Since they made no distinction between bright and dwarf galaxies, it is impossible to say whether the bright galaxy LFs are close to Gaussian or not. So, the absence of the dip in the field LF may be ascribed either to the same kind of effect present in Virgo, i.e. a large superposition between the bright and faint galaxy LFs, or to a real difference in shape between the bright galaxy LF of field and cluster galaxies. It is also possible that errors in the absolute magnitudes of field galaxies are so large that small features in the LF cannot be detected (Marzke et al. 1994 use magnitudes from the catalogue of Zwicky et al. 1961-1968, and assume an unperturbed Hubble flow to convert from apparent to absolute magnitudes).

## 6. Discussion and conclusions

We have shown that there is a bump in the bright part of the LF for galaxies in rich clusters (such as Coma, Shapley 8 and Abell 963) which is seen neither in the LF of a poor cluster (Virgo), nor in the LF of field galaxies. The presence of a bump in the bright part of the LF may therefore be related to the local environment.

ronments (*nature* point of view). In particular, the biased CDM theory predicts that brightest galaxies would be formed in the highest peaks of the initial density field (e.g. Frenk et al. 1985), and therefore would now be preferentially located in the central regions of the richest clusters. A systematic effect of luminosity segregation in the galaxy clustering is therefore expected, in particular for galaxies brighter than $M^*$, the magnitude corresponding to the knee of the LF (Valls-Gabaud et al. 1989). However, CDM with a value of the bias parameter much larger than unity is not consistent with the results from COBE (e.g. Wright et al. 1992; Myers et al. 1993).

Alternatively, the LF may be affected by environment-driven evolutionary effects in the galaxy properties (*nurture* point of view).

Enhanced star formation (generated by e.g. tidal effects) may increase the luminosity of galaxies in the central regions of the cluster. An early-type galaxy seen 1 Gyr after a burst of star formation would have 50 % of its visual luminosity produced by newly formed stars, and yet have an almost normal colour (Charlot & Silk 1994), so it would not deviate substantially from the average CM relation. However, such a recent starburst would enhance Balmer absorption lines, an effect not seen in the spectra of most galaxies in the central region of Coma (Caldwell et al. 1993).

Merging of galaxies may contribute to increase the number of bright early-type galaxies while decreasing the fraction of spirals. BCMs may be the extreme outcomes of this process. Following Mamon (1992), merging of galaxies in cluster environments is efficient at the present epoch, and roughly 60 % more merger events are expected in the central region of Coma than in Virgo (this estimate depends on the galaxy densities and velocity dispersions in the two clusters, see eq. 6 in Mamon 1992). The merger rate of bright galaxies can be increased if these are slowed down by dynamical friction, a process which should occur on a three times shorter timescale in Coma and Shapley 8 than in Virgo (see Table 3 of Biviano et al. 1992). In the merging scenario, BCMs would share a similar origin with other galaxies in the bright part of the LF; it is not surprising then that BCMs fit into the Erlang-shaped LF of bright galaxies.

It will be interesting to verify if galaxies belonging to the Erlang part of the LF are morphologically different from those in the Schechter part, and to compare them to the two populations of elliptical galaxies quoted by Capaccioli et al. (1992). Notice that for these authors one of these two populations has undergone merging, while the other has not. The photometric effect observed by Hamabe & Kormendy (1987) may be a signature of the first of these two populations of galaxies.

In a forthcoming paper, we will investigate this hypothesis, based on our photometric data in the V band,

to constrain the faint part of the Coma LF.

*Acknowledgements.* We are very grateful to Françoise Warin for her help in data reduction. We thank Gary Mamon, Didier Pelat, Armando Pisani, and Brigitte Rocca-Volmerange for useful discussions. We acknowledge financial support from GDR Cosmologie, CNRS, and from the EU Human Capital and Mobility program for one of us (AB).